%% file: A2744_SLanalysis.tex
\documentclass[useAMS,usenatbib]{mn2e}

\usepackage{graphicx}
\usepackage{txfonts}
\usepackage{amssymb,alltt}
\usepackage{color}
\usepackage{hyperref}
\usepackage{supertabular}

\hypersetup{colorlinks, linkcolor=blue}

\title[HFF Strong-Lensing Analysis of Abell 2744]{\textit{Hubble Frontier Fields}: A High-Precision Strong-Lensing Analysis of the Massive Galaxy Cluster Abell 2744 using $\sim$180 Multiple Images}
\author[Jauzac et al. 2014]
{M. Jauzac,$^{1,2}$\thanks{E-mail:
mathilde.jauzac@dur.ac.uk} J. Richard,$^{3}$ E. Jullo,$^4$ B. Cl\'ement,$^{3}$ M. Limousin,$^{4,5}$ J.-P. Kneib,$^{6,4}$ 
\newauthor
H. Ebeling,$^{7}$ P. Natarajan,$^{8}$ S. Rodney,$^{9,10}$ H. Atek,$^{6}$ R. Massey,$^{1}$ D. Eckert,$^{11}$ E. Egami,$^{12}$ 
\newauthor
M. Rexroth$^{6}$  
\\
\\
$^{1}$Institute for Computational Cosmology, Durham University, South Road, Durham DH1 3LE, U.K.\\
$^{2}$Astrophysics and Cosmology Research Unit, School of Mathematical Sciences, University of KwaZulu-Natal, Durban 4041, South Africa\\
$^{3}$CRAL, Observatoire de Lyon, Universit\'e Lyon 1, 9 Avenue Ch. Andr\'e, 69561 Saint Genis Laval Cedex, France\\
$^{4}$Laboratoire d'Astrophysique de Marseille - LAM, Universit\'e d'Aix-Marseille $\&$ CNRS, UMR7326, 38 rue F. Joliot-Curie, 13388 Marseille Cedex 13, France\\
$^{5}$Dark Cosmology Centre, Niels Bohr Institute, University of Copenhagen, Juliane Maries Vej 30, DK-2100 Copenhagen, Denmark\\
$^{6}$Laboratoire d'Astrophysique, Ecole Polytechnique F\'ed\'orale de Lausanne (EPFL), Observatoire de Sauverny, CH-1290 Versoix, Switzerland\\
$^{7}$Institute for Astronomy, University of Hawaii, 2680 Woodlawn Drive, Honolulu, Hawaii 96822, USA\\
$^{8}$Department of Astronomy, Yale University, 260 Whitney Avenue, New Haven, CT 06511, USA\\
$^{9}$Department of Physics and Astronomy, The Johns Hopkins University, 3400 N. Charles St., Baltimore, MD 21218, USA\\
$^{10}$Hubble Fellow\\
$^{11}$Astronomy Department, University of Geneva, 16 ch. d'Ecogia, CH-1290 Versoix, Switzerland\\
$^{12}$Steward Observatory, University of Arizona, 933 North Cherry Avenue, Tucson, AZ, 85721, USA} 

\begin{document}

\date{Accepted 2015 June 23. Received 2015 June 23; in original form 2014 October 02}

\pagerange{\pageref{firstpage}--\pageref{lastpage}} \pubyear{2015}

\maketitle

\label{firstpage}

\begin{abstract}
\input{abstract}
\end{abstract}



\section{Introduction}
\label{intro}
\input{introduction}

\section{\textit{Hubble Frontier Fields} Observations}
\label{observations}
\input{HFFdata}

\input{table1.tex}
\section{Strong Lensing Analysis}
\label{SLanalysis}

\begin{figure*}
\begin{center}
\includegraphics[width=0.985\textwidth]{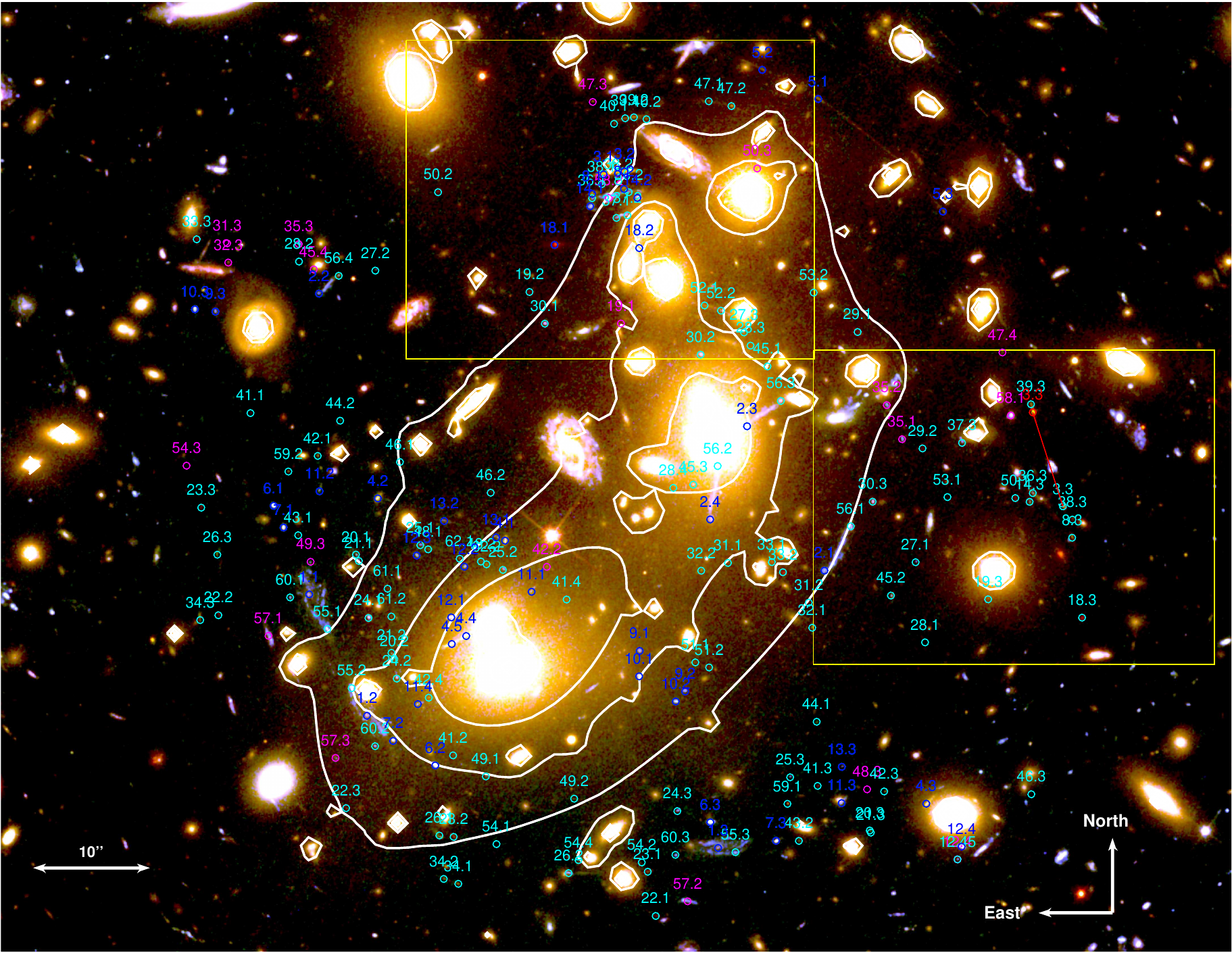}\\
\includegraphics[width=0.49\textwidth]{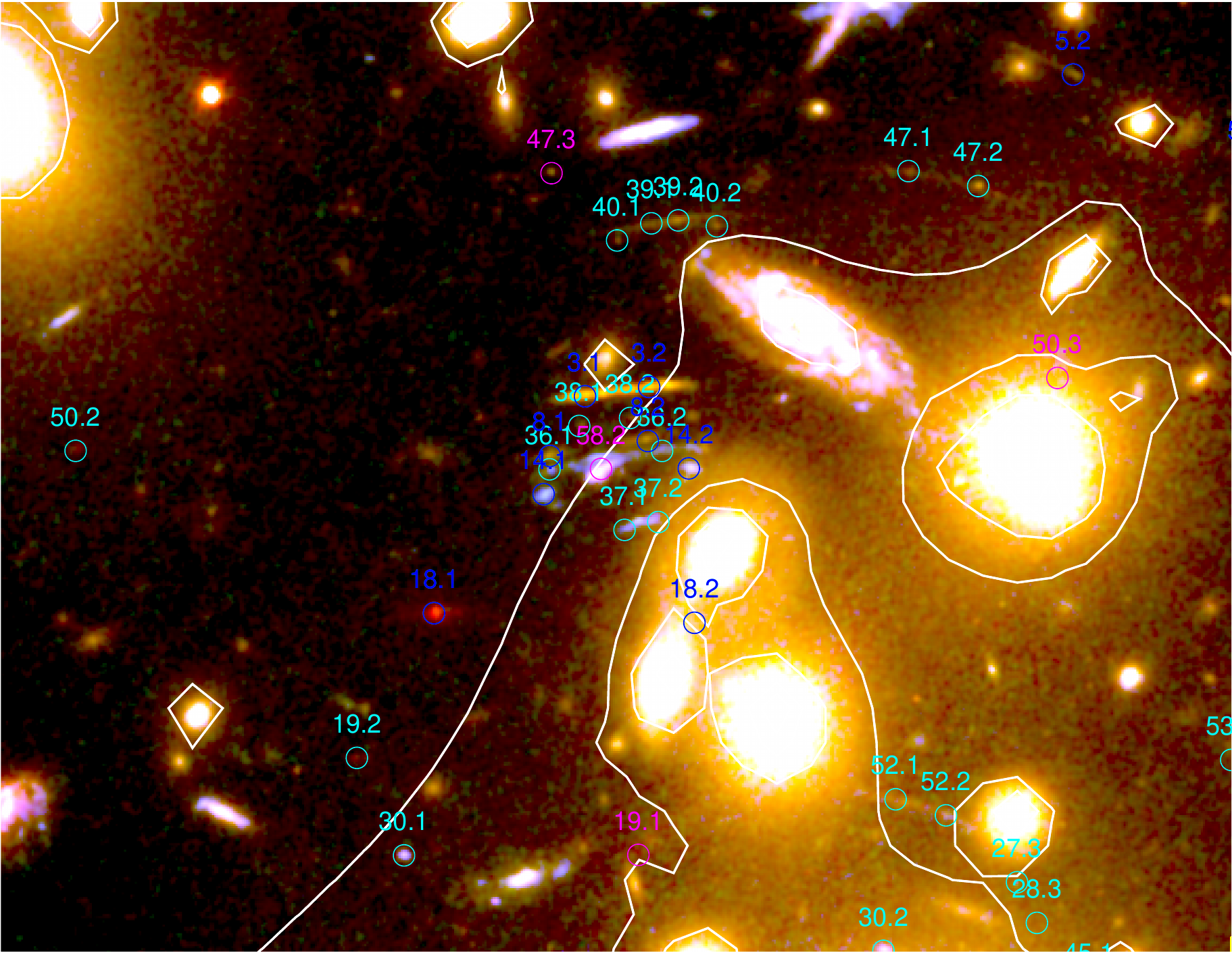}
\includegraphics[width=0.49\textwidth]{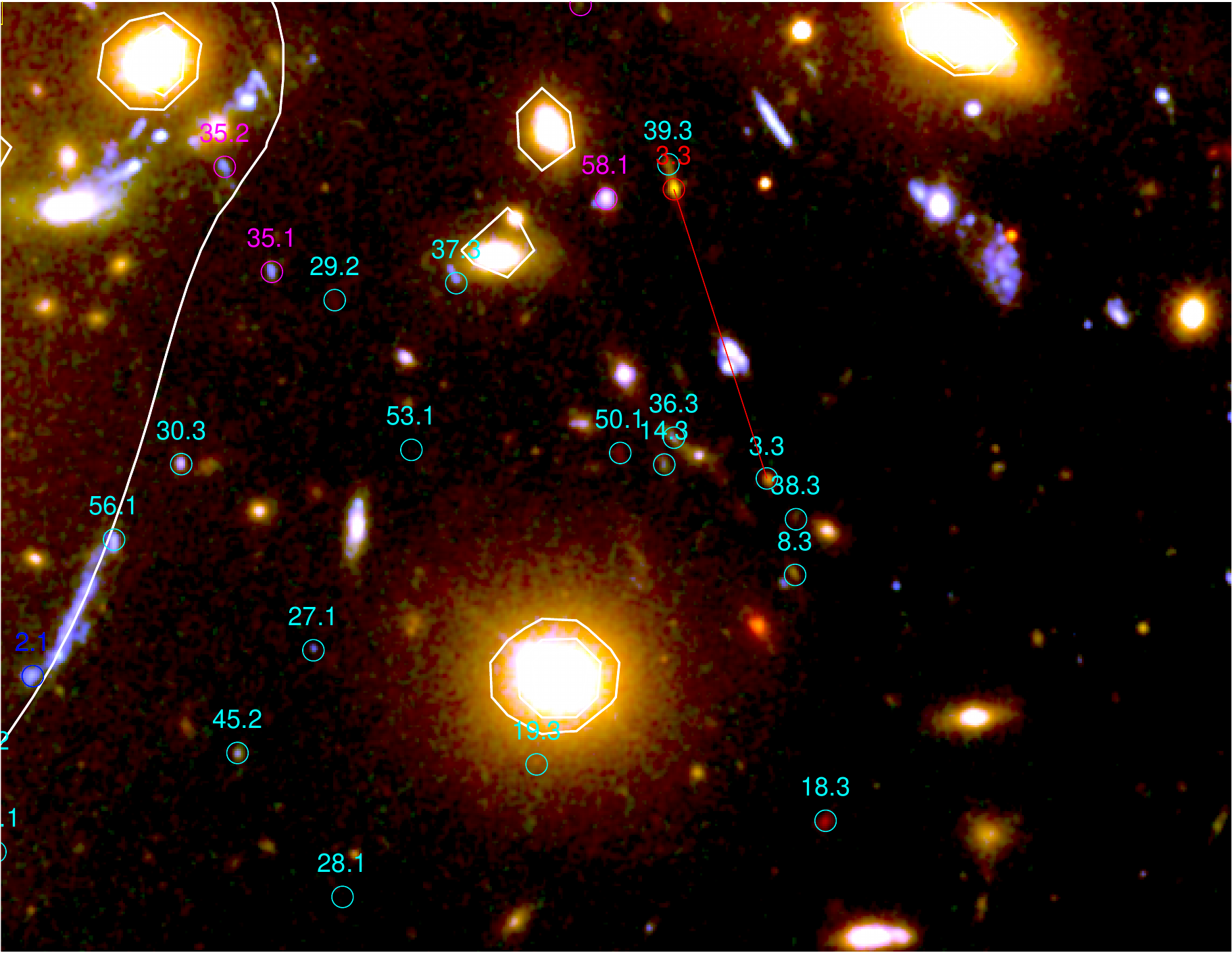}\\
\caption{Overview of all multiple-image systems used in this study. The pre-HFF systems are shown in dark blue. The most secure HFF identifications, used to optimise the lens model in the \textit{image plane} (152 images) are shown in cyan; the less secure candidates (7 images) are shown in magenta. The underlying colour image is a composite created from HST/ACS images in the F814W, F606W, and F435W passbands. Mass contours of the best-fit strong-lensing model are shown in white. 
The zoomed stamps show the particular configuration of the multiply imaged systems in the northern part of the cluster core (systems 3, 8, 14, 36, 37, 38, 39, and 40). In the right panel, one can see highlighted with a red line the shift of position between the old identification of 3.3 and the new one.}
\label{multiples}
\end{center}
\end{figure*}

\subsection{Methodology}
\input{SLmethod}

\input{table2.tex}

\subsection{Multiple-Image Systems}
\input{SLconstraints}

\label{slcons}

\section{Strong-Lensing Mass Measurement}
\label{SLmass}

\begin{center}
\begin{figure}
\hspace*{-3mm}\includegraphics[width=0.5\textwidth,angle=0.0]{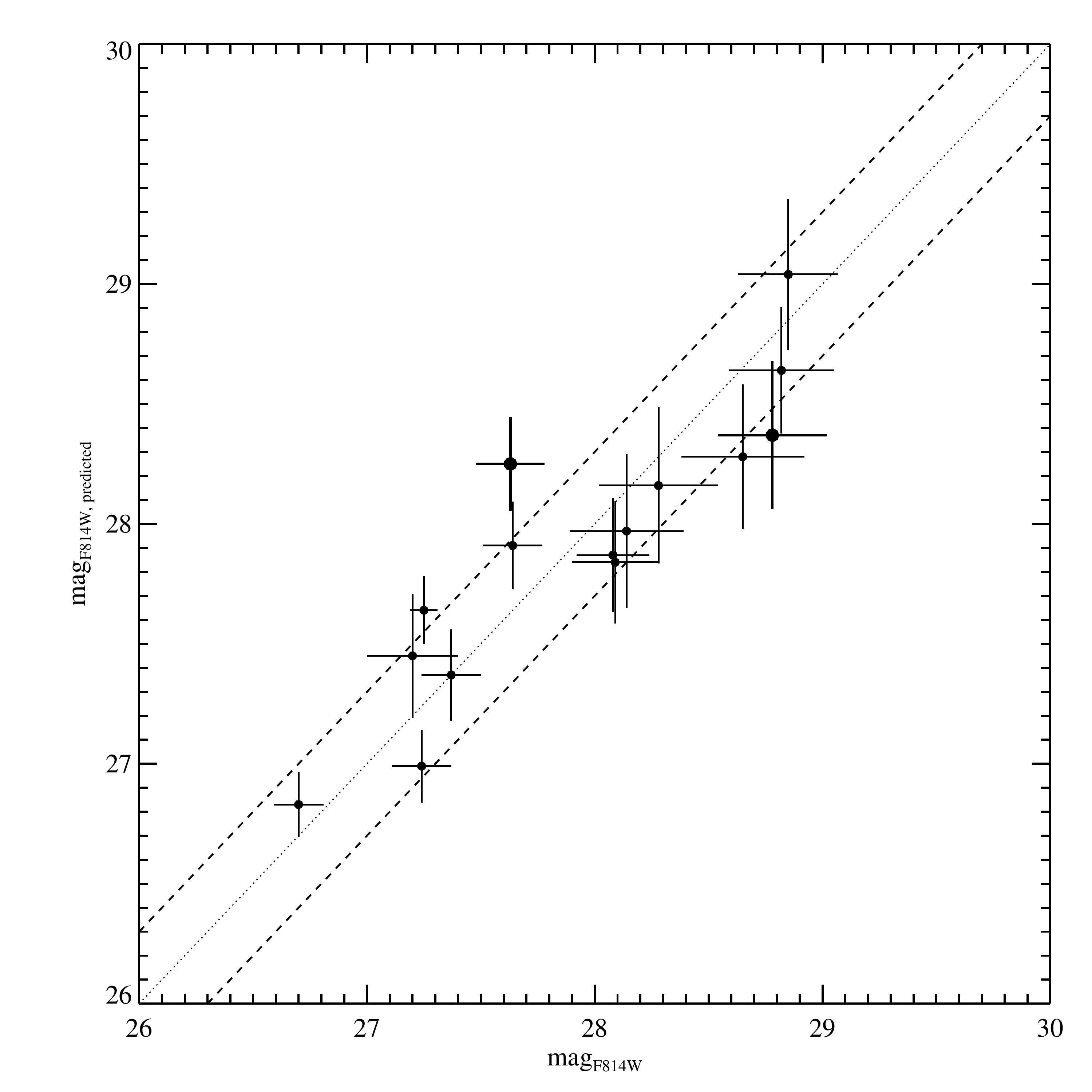}
\caption{Magnitudes in the $F814W$-band predicted by our model against the observed ones for some of the multiple images. The solid line shows the one-to-one relation, while the dashed lines represent a dispersion of 0.3 $mag$.
}
\label{maglens_delens}
\end{figure}
\end{center}

\input{SLmass}

\begin{center}
\begin{figure}
\hspace*{-3mm}\includegraphics[width=0.5\textwidth,angle=0.0]{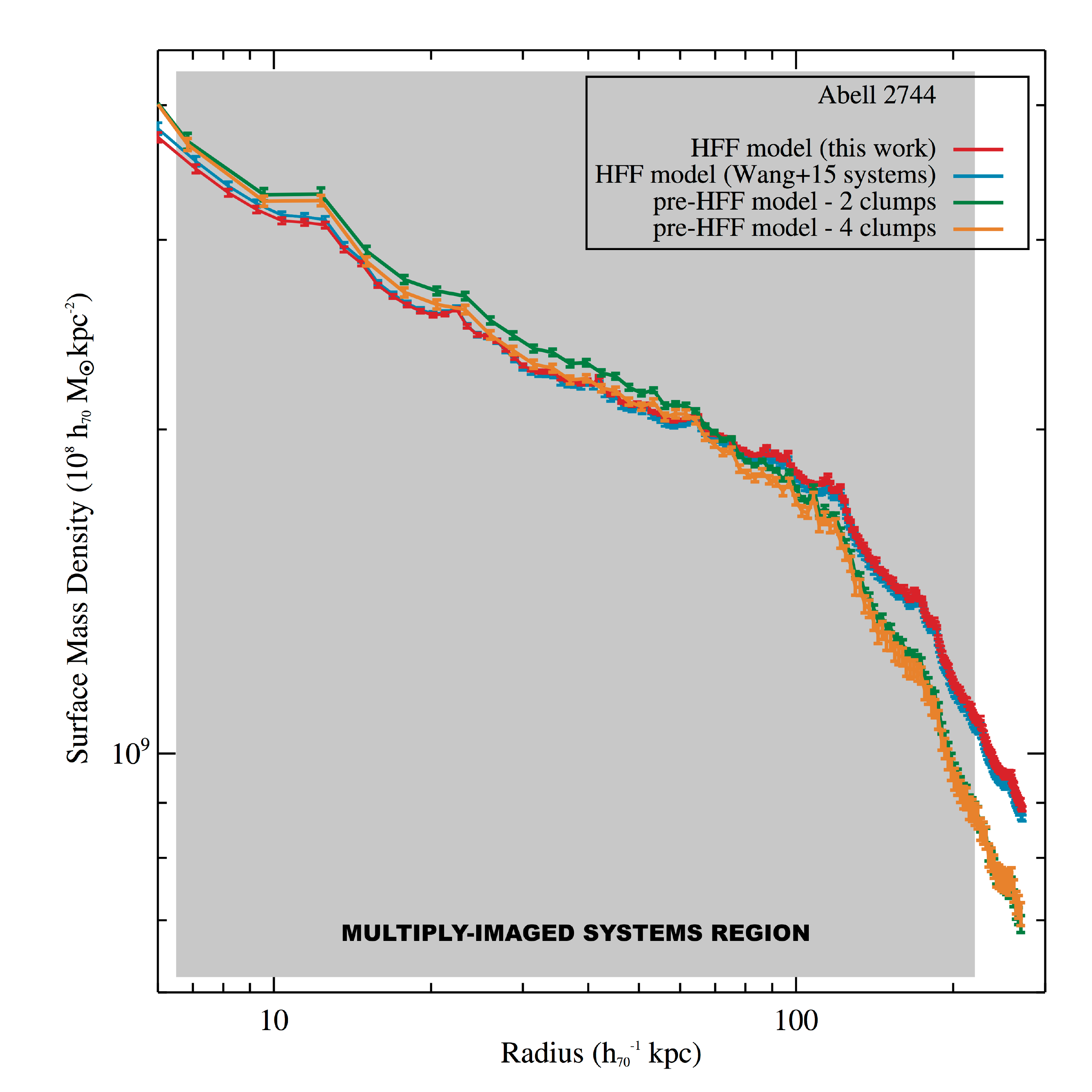}
\caption{Surface mass density profile obtained with the new HFF constraints (red), as well as the HFF model using the \citet{wang15} systems set (cyan). Also shown are the profiles obtained with the 2- and 4-component pre-HFF mass models (green and orange, respectively). The grey shaded region highlight the region which includes multiply-imaged systems.
}
\label{densprof}
\end{figure}
\end{center}

\section{Discussion}
\label{discussion}

\begin{center}
\begin{figure*}
\includegraphics[width=0.46\textwidth,angle=0.0]{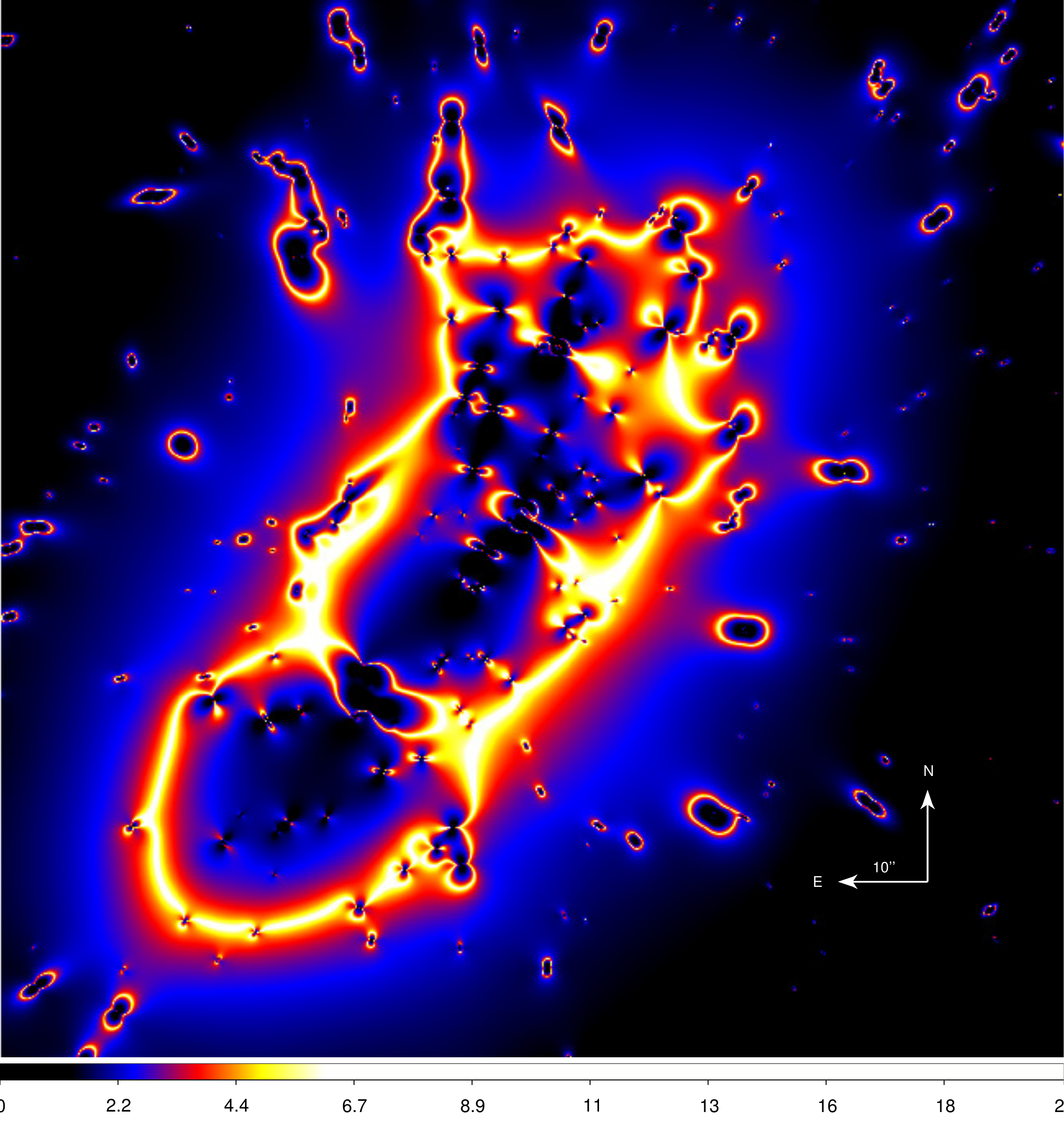}\ 
\hspace*{3mm}\includegraphics[width=0.5\textwidth,angle=0.0]{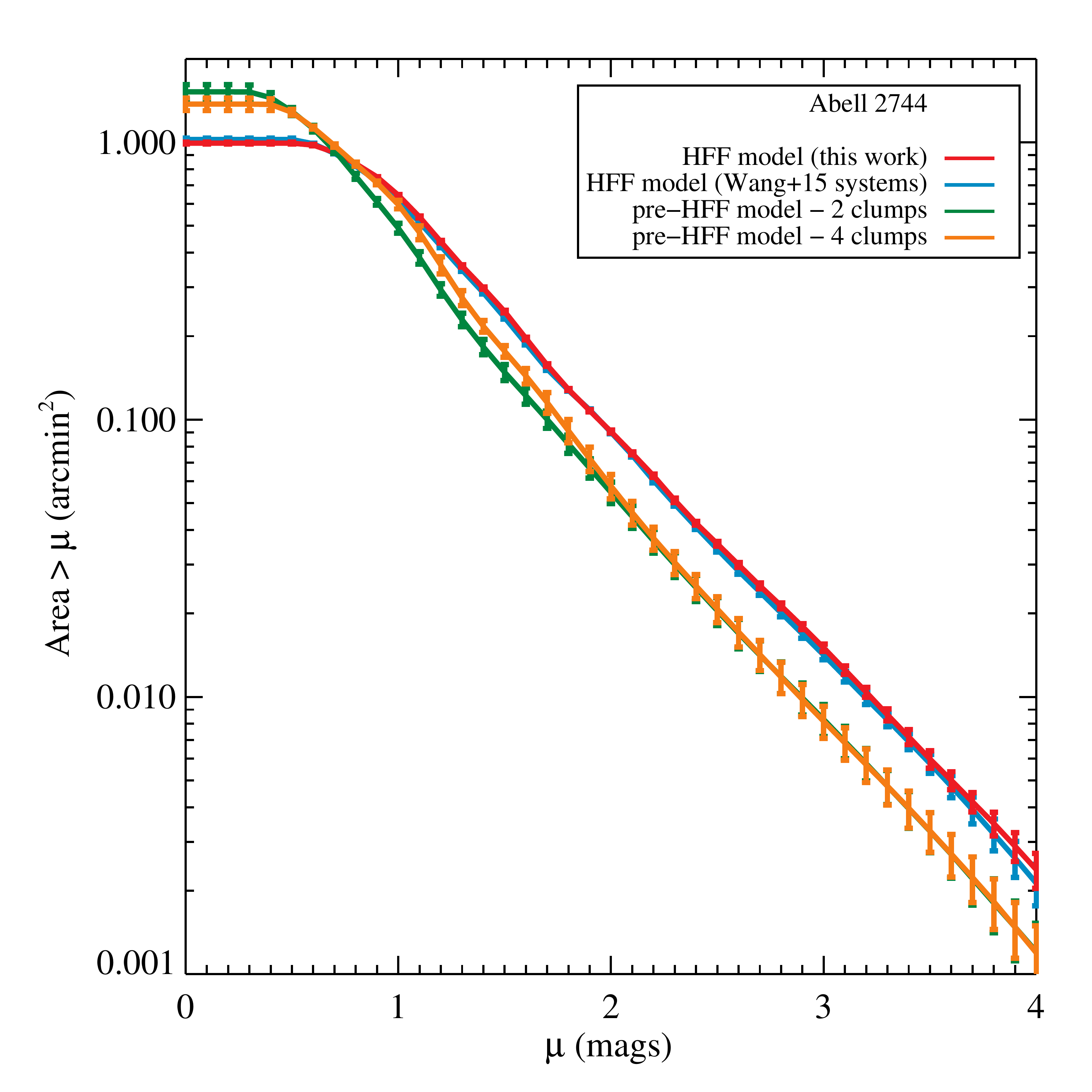}\ 
\\
\caption{\textit{Left panel: }Magnification map obtained from our HFF lens model for a source at $z_{\rm S}=9$. \textit{Right panel: }Surface area in the source plane covered by WFC3 at a magnification above a given threshold $\mu$ for the present HFF model (red), the HFF model using the set of multiply-imaged systems defined by \citet{wang15} (cyan), the pre-HFF 2 clumps model (green), and the pre-HFF 4 clumps model (orange). 
}
\label{diffampli}
\end{figure*}
\end{center}

\input{discussion_hff}

\section*{Acknowledgments}
We would like to thank the anonymous referee for his/her constructive comments and suggestions.
JR acknowledges support from the ERC starting grant CALENDS and the CIG grant 294074. This work was supported by the Science and Technology Facilities Council [grant number ST/F001166/1] and used the DiRAC Data Centric system at Durham University, operated by the Institute for Computational Cosmology on behalf of the STFC DiRAC HPC Facility (www.dirac.ac.uk [www.dirac.ac.uk]). This equipment was funded by BIS National E-infrastructure capital grant ST/K00042X/1, STFC capital grant ST/H008519/1, and STFC DiRAC Operations grant ST/K003267/1 and Durham University. DiRAC is part of the National E-Infrastructure. MJ, EJ, and ML acknowledge the M\'esocentre d'Aix-Marseille Universit\'e (project number: 15b030).  This study also benefited from the facilities offered by CeSAM (CEntre de donn\'eeS Astrophysique de Marseille ({\tt http://lam.oamp.fr/cesam/}). ML acknowledges the Centre National de la Recherche Scientifique (CNRS) for its support, and the Dark cosmology centre, funded by the Danish National Research Fundation. EJ and ML acknowledge the Centre National d'Etude Spatial (CNES) for its support. JPK and HA acknowledge support from the ERC advanced grant LIDA. PN acknowledges support from the National Science Foundation via the grant AST-1044455, AST-1044455, and a theory grant from the Space Telescope Science Institute HST-AR-12144.01-A. RM is supported by the Royal Society. Support for SAR was provided by NASA through Hubble Fellowship grant \#HST-HF-51312.01 awarded by the Space Telescope Science Institute, which is operated by the Association of Universities for Research in Astronomy, Inc., for NASA, under contract NAS 5-26555.
Based on observations made with the NASA/ESA Hubble Space Telescope, obtained from the data archive at the Space Telescope Science Institute. STScI is operated by the Association of Universities for Research in Astronomy, Inc. under NASA contract NAS 5-26555.

\bibliographystyle{mn2e}
\bibliography{reference}

\input{table3.tex}



\label{lastpage}

\end{document}

%% file: abstract.tex
We present a high-precision mass model of galaxy cluster Abell 2744, based on a strong-gravitational-lensing analysis of  the \emph{Hubble Space Telescope Frontier Fields} (HFF) imaging data, which now include both \emph{Advanced Camera for Surveys}  and \emph{Wide-Field Camera 3} observations to the final depth.
Taking advantage of the unprecedented depth of the visible and near-infrared data, we identify 34 new multiply imaged galaxies, bringing the total to 61, comprising 181 individual lensed images.  In the process, we correct previous erroneous identifications and positions of multiple systems in the northern part of the cluster core.
With the \textsc{Lenstool} software and the new sets of multiple images, we model the cluster using two cluster-scale dark matter halos plus galaxy-scale halos for the cluster members. Our best-fit model predicts image positions with an \emph{RMS} error of 0.69$\arcsec$, which constitutes an improvement by almost a factor of two over previous parametric models of this cluster.
We measure the total projected mass inside a 200~kpc aperture as ($2.162\pm 0.005$)$\times 10^{14}M_{\odot}$, thus reaching 1\% level precision for the second time, following the recent HFF measurement of MACSJ0416.1-2403. Importantly, the higher quality of the mass model translates into an overall improvement by a factor of 4 of the derived magnification factor. 
Together with our previous HFF gravitational lensing analysis, this work demonstrates that the HFF data enables high-precision mass measurements for massive galaxy clusters and the derivation of robust magnification maps to probe the early Universe.

%% file: introduction.tex
Since the end of the 1980s and the first observational confirmation of the lensing hypothesis in Abell 370 \citep[][]{soucail88}, gravitational lensing has been recognised as a powerful tool to map the mass distribution in galaxy clusters. The bending of light from distant galaxies by foreground clusters allows astronomers to  $i)$ directly measure the total (dark and baryonic) matter distribution, $ii)$ image very distant galaxies using galaxy clusters as {`cosmic telescopes'}, and $iii)$ constrain the geometry of the Universe (for reviews, see e.g.\ \citealt{2010RPPh...73h6901M} and \citealt{KN11,hoekstra13}). 
In the past few decades, the unparalleled power of the \textit{Hubble Space Telescope} (HST) has transformed this area of research. HST's high angular resolution and sensitivity combined with colour information from imaging through several filters allow a robust and efficient identification of lensed galaxies, as demonstrated in many in-depth studies \citep[for Abell 1689, one of the best studied cluster to date, using the \emph{Advanced camera for Survey} see e.g. ][]{coe10,limousin07b,broadhurst05a}.  

The leading role of HST for lens studies has been recognised by the community through numerous general observing programs, and specifically through two recent large allocations: the multi-cycle treasury CLASH project \citet{postman12} and the novel \textit{Hubble Frontier Fields}\footnote{http://www.stsci.edu/hst/campaigns/frontier-fields/} (HFF) project.
 With a total of 140 HST orbits for each of six massive cluster lenses, the goal of the HFF is to  probe the distant and early Universe to an unprecedented depth of $mag_{\mathrm{AB}}\!\sim\!29$ in seven passbands (3 with ACS, 4 with WFC3).
In a coordinated multi-team effort, mass models\footnote{http://archive.stsci.edu/prepds/frontier/lensmodels/} of all six HFF cluster lenses were derived from pre-HFF data to provide the community with a first set of magnification maps \citep[see in particular][]{johnson14,coe15,richard14}.
The first two clusters observed as part of the HFF initiative are MACSJ0416.1$-$2403 \citep{ME12} and Abell 2744 \citep{abell89}; to date lensing studies based on these data have been conducted by \cite{jauzac14a,jauzac15,diego14,grillo14,zitrin14,wang15,ogrean15}.

In this paper, we focus on improving earlier strong-lensing analyses of Abell 2744 (also known as AC118 and MACSJ0014.3$-$3022), a very X-ray luminous cluster at $z{=}0.308$, featuring $L_{X} =3.1\times 10^{45}$erg s$^{-1}$ in the 2--10 keV band \citep[][]{allen98} and $1.4\times 10^{45}$ erg s$^{-1}$ in the 0.1--2.4  keV band \citep{ebeling10}. 
This system has been extensively studied, and was first identified as an active major merger by \cite{giovannini99} and \cite{govoni01a} based on the detection of a powerful and extended radio halo.  Subsequent X-ray studies \citep[][]{kempner04,zhang04,owers11} confirmed this hypothesis and detected numerous substructures within the cluster field.  Kinematic studies by  \cite{girardi01}, \cite{boschin06}, and \cite{braglia07} revealed a bimodal velocity dispersion in the cluster centre, as well as a third group of cluster members in the North-West, close to one of the X-ray peaks. \cite{shan10} studied a sample of 38 clusters with X-ray observations as well as high-resolution lensing observations coming from the HST/\emph{Wide Field and Planetary Camera 2} (WFPC2) archive. In their analysis, Abell 2744 was identified as the cluster with the largest offset between X-ray and lensing signals.

The first gravitational-lensing analysis of Abell 2744 was conducted by \cite{smail97a}, who identified strong lensing features and also performed a weak-lensing analysis based on HST/WFPC-2 data. \cite{allen98} studied the discrepancy between the X-ray and strong-lensing masses of Abell 2744, which supports the hypothesis that the system is an active merger.  Results of a weak-lensing  analysis of Abell 2744 were also presented by \cite{cypriano04} as part of a study of a larger cluster sample. The first investigation of Abell 2744 to combine strong-lensing, weak-lensing, and X-ray analyses was conducted by \cite{merten11} using HST/ACS\footnote{Prop. ID: 11689, P.I.: R. Duke}, VLT, Subaru imaging, and \textit{Chandra} observations. More recently, \cite{lam14} performed a strong-lensing analysis of Abell 2744, based on pre-HFF data. \cite{wang15} presented an HFF strong-lensing analysis which includes 72 multiple images, selected using some specific criteria combining the colors and morphology in each system. In \cite{atek14a,atek15}, \cite{laporte14} and \cite{ishigaki15}, the first HFF $z>6$ galaxy candidates lensed by Abell 2744 are presented, with a candidate at $z\sim$10 \citep{zitrin14}.

In this paper we present results from a new and improved strong-lensing analysis of the complete HFF ACS and WFC3 observations of Abell 2744.  We adopt the $\Lambda$CDM concordance cosmology with $\Omega_{m} = 0.3$, $\Omega_{\Lambda} = 0.7$, and a Hubble constant $H_0 = 70$~km$\,$s$^{-1}\,$Mpc$^{-1}$. Magnitudes are quoted in the AB system.

%% file: HFFdata.tex
The HFF observations of Abell 2744 (ID: 13495, P.I: J. Lotz) were obtained with WFC3 between October 25$^{th}$ and November 28$^{th}$ 2013 in four filters (F105W, F125W, F140W, and F160W) for total integration times corresponding to 24.5, 12, 10, and 14.5 orbits respectively. The HFF ACS observations  were obtained more recently, between May 14$^{th}$ and July 1$^{st}$ 2014, in three filters (F435W, F606W, and F814W) for total integration times corresponding to 24, 14, and 46 orbits, respectively.  We use the self-calibrated data (version v1.0) with a pixel size of 0.03\arcsec, provided by STScI\footnote{http://archive.stsci.edu/pub/hlsp/frontier/a2744/images/hst/}.

%% file: table1.tex

\begin{table}
\begin{center}
\begin{tabular}[h!]{cccc}
\hline
\hline
\noalign{\smallskip}
Component  & \#1 & \#2 & L$^*$ elliptical galaxy \\
\hline
$\Delta$ \textsc{ra}  & $-4.9^{+0.2}_{-0.1}$  &  $-15.7^{+0.1}_{-0.2}$ & --  \\
$\Delta$ \textsc{dec} & 2.7 $^{+0.3}_{-0.4}$  & $-17.2^{+0.2}_{-0.1}$ & --  \\
$e$ & 0.28 $\pm$0.008 & 0.61 $\pm$ 0.004 & -- \\
$\theta$ & 65.0$^{+0.5}_{-0.5}$  & 43.3$^{+0.6}_{-0.4}$ & -- \\
r$_{\mathrm{core}}$ (\footnotesize{kpc}) & 214.4$^{+2.2}_{-1.7}$  & 43.5$^{+0.9}_{-0.9}$  & [0.15] \\
r$_{\mathrm{cut}}$ (\footnotesize{kpc}) & [1000] & [1000] & 22.2$^{+1.9}_{-1.5}$ \\
$\sigma$ (\footnotesize{km\,s$^{-1}$}) &  1263$^{+6}_{-6}$ & 134$^{+7}_{-13}$ & 154.9$\pm$ 2.4 \\
\noalign{\smallskip}
\hline
\hline
\end{tabular}
\caption{Best-fit PIEMD parameters for the two large-scale dark-matter halos, as well as for the L$^{*}$ elliptical galaxy. 
Coordinates are quoted in arcseconds with respect to $\alpha=3.586259, \delta=-30.400174$.
Error bars correspond to the $1\sigma$ confidence level. Parameters in brackets are not optimised.
The reference magnitude for scaling relations is $mag_{\rm{F814W}} = 19.44$.
}
\label{tableres}
\end{center}
\end{table}

%% file: SLmethod.tex
We here provide only a brief summary of our method and refer the reader to \cite{kneib96}, \cite{smith05}, \cite{verdugo11}, \cite{richard11b} for detailed explanations.  Our mass model is composed of large-scale dark-matter haloes, whose individual masses are larger than a typical galaxy group (of order of 10$^{14}$\,M$_{\sun}$ within 50$\arcsec$), plus small-scale mass halos associated with individual cluster members, usually large elliptical galaxies.  As in our previous work, we model all mass components as dual Pseudo Isothermal Elliptical Mass Distributions \citep[dPIE,][]{limousin05,eliasdottir07,jauzac14a}, characterised by velocity dispersion $\sigma$, core radius $r_{\rm core}$, and scale radius $r_s$. 
Note that contrary to non-parametric approaches, such as those presented in \cite{jauzac12,jauzac15}, the optimisation we use here does not allow for halos to be set to a mass value of zero (i.e. the velocity dispersion cannot reach a value of zero). Therefore, we rely on the $\chi^2$ and RMS statistics to rank different models and priors with respect to the observed positions of multiply-imaged galaxies.

For mass perturbations associated with individual cluster galaxies, we fix the geometrical dPIE parameters (centre, ellipticity, and position angle) at the values measured from the cluster light distribution \citep[see, \emph{e.g.}][]{kneib96,limousin07b,richard10a} and use empirical scaling relations to relate the dynamical dPIE parameters (velocity dispersion and scale radius) to the galaxies' observed luminosity \citep[][]{richard14}.  For an $L_{\ast}$ galaxy, we optimise the velocity dispersion between 100 and 250 km\,s$^{-1}$ and force the scale radius to less than 70 kpc to account for the tidal stripping of galactic dark-matter haloes \citep{limousin07a,limousin09a,natarajan09,wetzel10}.

%% file: table2.tex
\begin{table}
\begin{center}
\begin{tabular}[h!]{ccccc}
\hline
\hline
\noalign{\smallskip}
Models & $\chi^{2}$ & RMS & dof & $\mu_{13mul}$ \\
\hline
pre-HFF 5 clumps & 4.99 & 1.26$\arcsec$ & 25 & 6.04$\pm$0.49 \\
pre-HFF 4 clumps & 4.47 & 0.84$\arcsec$ & 26 & 4.56$\pm$0.14 \\
pre-HFF 2 clumps & 3.23 & 0.79$\arcsec$ & 36 & 4.69$\pm$0.32 \\
HFF & 2.77 & 0.79$\arcsec$ & 138 & 5.61$\pm$0.10\\
\noalign{\smallskip}
\hline
\hline
\end{tabular}
\caption{
Parameters obtained for the three different models used in this analysis to compare our HFF model with pre-HFF ones. The \emph{dof} gives the number of degrees of freedom in each model, and the $\mu_{13mul}$ gives the average magnification obtained using a set of 13 multiple images common to all models.
The \emph{`pre-HFF 5 clumps'} is \citet{richard14} model; the \emph{`pre-HFF 4 clumps'} is \citet{richard14} model with the identification of counter-image in system 3 corrected; the \emph{`pre-HFF 2 clumps'} is \citet{richard14} model with the identification of counter-image in system 3 corrected, and without the N and NW substructures; the \emph{`HFF'}s model is the one presented in this paper.
}
\label{tablemodel}
\end{center}
\end{table}

%% file: SLconstraints.tex
\subsubsection{HFF multiply-imaged systems}
The secure identification of multiple-image systems is key to building a robust model of the mass distribution within the cluster lens. The first detailed strong-lensing analysis, using pre-HFF data of Abell 2744,  identified 34 images of 11 background galaxies in the redshift range $2\!<\!z\!<\!4$ \citep{merten11}. Based on the same data, but in a community-wide effort for the HFF mass model initiative, the number of secure systems increased to 18, comprising 55 images. 
Three of these systems have been spectroscopically confirmed, with redshifts of 3.98, 3.58 and 2.019 for systems 3, 4, and 6, respectively \citep[see ][]{richard14,johnson14}. 

For the present study, we scrutinised the new, deep HFF ACS and WFC3 images, using the predictive power of the \citet[][hereafter R14]{richard14} mass model and report an even larger set of multiple images. To this end, we computed the cluster's gravitational-lensing deflection field from the image plane to the source plane, on a grid with a spacing of 0.2 arcsec/pixel. Since the transformation scales with redshift as described by the distance ratio $D_{LS}/D_{OS}$, where $D_{LS}$ and $D_{OS}$ are the distances between the lens and the source, and the observer and the source respectively, it needs to be computed only once enabling efficient lens inversion. We also determined the critical region at redshift $z=7$ as the area within which to search for multiple images in the ACS data.  A thorough visual inspection of all faint galaxy images in this region, combined with an extensive search for plausible counter images, revealed a total of 34 new multiply imaged systems, bringing the total of multiple images identified in Abell 2744 to 181 over 61 systems (Fig.~\ref{multiples} and Table~\ref{multipletable}).
More recently, \cite{wang15} presented an HFF strong- and weak-lensing analysis of Abell 2744, in which they provide spectroscopic redshifts for systems 1 and 2, at z=1.50 and z=1.20 respectively, derived from the \emph{Grism Lens- Amplified Survey from Space} (GLASS) observations. They also present new systems, namely 55, 56 detected by \cite{lam14} and which represent the other extremities of systems 1 and 2 respectively; 57, 58, 59, all three are high-redshift lensed candidates detected by \cite{ishigaki15}, and finally system 60 which is a new system. We have included in our mass model systems 55 and 56 with their respective spectroscopic redshifts, as well as systems 59 and 60. However we are not convinced by systems 57 and 58, thus we included them as candidates. We did not retain the GLASS spectroscopic redshift for system 2, of lower quality, because it is inconsistent with the previous photometric redshift and spectroscopic redshift estimates \citep{johnson14}.

Table~\ref{multipletable} lists the coordinates, as well as the redshifts (predicted by our model, $z_{model}$, or spectroscopic, $z_{spec}$, if available), the F814W-band magnitudes, $mag_{F814W}$, and their magnification (measured with our best-fit mass model). The magnitudes were measured using \textsc{Sextractor} \citep{BA96}.  Note that for some of the images, reliable measurements were rendered impossible due to light contamination from much brighter objects.
Systems 15, 16, and 17 around the North and North-West sub-structures are not included in either Fig.~\ref{multiples} or Table~\ref{multipletable}, as we do not use them in our optimisation, but we refer the reader to R14, \cite{johnson14}, or \cite{lam14} for their coordinates.

For the modelling of the cluster lens, we adopt a conservative approach and use only the 54 most securely identified systems comprising 154 individual images; we propose the remaining identifications as candidates for multiply imaged systems.  We consider a system secure if it meets all of the following criteria: the different images have (1) similar colours, (2) show morphological similarities (for resolved images), and finally  (3) a plausible geometrical configuration. As for MACSJ0416.1$-$2403 \citep{jauzac14a}, the total number of multiply-imaged sets used in the optimisation has increased by a factor of 3, and the precision of the lens model in the core region of maximal magnification improves dramatically.

\subsubsection{Multiplicity of high-redshift candidates} 

\noindent We also confirm and include systems proposed to lie at $z>5$ by \cite{atek14a} and \cite{atek15} but identify object 22.1  as a more convincing counter image in terms of position and colour for System 4 in \citet{atek14a}. Their image 4.1 is now predicted to be a single image. We note also that image 18.3 has a measured spectroscopic redshift of 5.66 (Cl\'ement et al., \emph{in prep.}). 
We also include the $z\sim 10$ system identified by \cite{zitrin14}, which is well reproduced by our model.

\subsubsection{Revisiting Northern multiply imaged systems}

\noindent The deeper ACS images of Abell 2744 revealed several new multiple systems to the North of its brightest cluster galaxy (BCG), a region within which all previous strong-lensing analyses failed to identify correct counter images (such as 8.3, 14.3, and 18.3), or could not reproduce their positions to better than 2$\arcsec$ \citep[R14;][]{johnson14,lam14}.
The newly discovered multiple systems call into question the identification of image 3.3 by \cite{johnson14}, \cite{lam14}, and R14. By performing our strong-lensing analysis without this system, we predict the location of image 3.3 to lie 8$\arcsec$ south of the previous identification used by all modellers, which is now assigned to a different system to the North. The reason for this misidentification is due to the similarity in colors, but also in the lens reconstruction, as shown by \cite{lam14}. However, with this corrected position, we manage to identify all previously missing counter-images, and all systems in this region are now reproduced to better than 0.3$\arcsec$. 
To support our statements, we have run different tests (i.e., different models including the previous position for 3.3, the new 3.3 position and none of them) that are presented in the following section, all agreeing with the fact that the previous identification of 3.3 was wrong. However, we agree with \cite{lam14} that the colours are well matched by the old image 3.3. Images 3.1 and 3.2 represent a pair only lensing part of a source galaxy, while 3.3 is the counter-image lensing the entire source. In such a configuration, it is possible to observe different colours, due to intrinsic colour gradients within the source galaxy.

%% file: SLmass.tex
The starting point for our modelling process is the distribution of cluster galaxies. As described in \cite{merten11}, Abell 2744 is a highly complex system, with one main component in the SE and three p-scale substructures in the North, in the North-West, and in the West, labelled as \emph{Core}, \emph{N}, \emph{NW}, and \emph{W} in their paper. All these substructures host overdensities of bright cluster ellipticals: the core region is dominated by three brightest cluster galaxies (BCGs); the N, NW, and W substructures each host one BCG.  Following this optical morphology, the pre-HFF model of R14  thus included one cluster-scale dark-matter halo at the location of each of the five BCGs that define the centres of the overall large-scale distribution of light from all cluster galaxies. The \emph{W} substructure was not included in the mass model because it is outside the high-resolution HST imaging, so no strong-lensing features could be identified. The resolution of the ground-based imaging (VLT/FORS1) does not allow for any identification. 

For our revised model based on the new HFF data, we simplified this mass model as follows: i) since no spectroscopic redshifts are currently available for multiple images around the N and NW substructures, we removed the corresponding mass halos, and a discussion about the impact of this removal is presented in Section~\ref{discussion}; ii) the core of the cluster is now modeled using only two cluster-scale halos instead of three, for reasons explained later in this Section. Consequently, our model contains only two cluster-scale halos.
During the optimisation process, the position of these large-scale halos is allowed to vary within 20$\arcsec$ of the associated light peak. In addition, we limit the ellipticity, defined as $e=(a^2+b^2)/(a^2-b^2)$, to values below 0.7,  while the core radius and the velocity dispersion are allowed to vary between 1$\arcsec$ and 30$\arcsec$, and 300 and 1\,000 km\,s$^{-1}$, respectively. The scale radius, by contrast, is fixed at 1\,000\,kpc, since strong-lensing data alone do not probe the mass distribution on such large scales. In addition to the two cluster-scale dark-matter halos, we also include perturbations by 733 probable cluster members (R14) by associating a galaxy-scale halo to each of them. Finally, we add two galaxy-scale halos to specifically model two of the three BCGs in the core region, as there were multiple images in their immediate proximity. Using the set of the most securely identified multiply imaged galaxies described in Sect.~\ref{SLanalysis} and shown in Fig.~\ref{multiples}, we optimise the free parameters of this mass model using the publicly available \textsc{Lenstool} software\footnote{http://projects.lam.fr/repos/lenstool/wiki}.

The best-fit model optimised in the \emph{image plane} predicts image positions that agree with the observed positions to within an RMS of {\color{red}0.79$\arcsec$}.  For MACSJ0416.1$-$2403, we found an RMS of 0.68$\arcsec$ \citep{jauzac14a}, for a total of 68 multiply imaged systems.  
These remarkably low RMS values obtained for the first two HFF clusters improve dramatically on the ones obtained in previous models from the literature that adopt a similar \emph{a priori} assumption of light tracing mass  \citep[e.g. ][for Abell 1689]{broadhurst05b,halkola06,limousin07b}.  The RMS value obtained for Abell 2744 represents an improvement of a factor of $\sim$1.5 over previous lensing mass reconstructions of this cluster.  \cite{johnson14} quote an RMS value of 0.4$\arcsec$ for their pre-HFF model using 15 multiply imaged systems (64 images). In R14, our pre-HFF best-fit mass model yields 1.26$\arcsec$ using 18 multiply imaged systems (55 images) combined with weak-lensing constraints. The parameters describing our best-fit mass model are listed in Table~\ref{tableres}; contours of the mass distribution are shown in Fig.~\ref{multiples}. Although allowed to vary within 20$\arcsec$ of their associated light peak, the final positions of the two cluster-scale halos predicted by the model coincide much more closely with the light peaks. Fig.~\ref{maglens_delens} presents the observed magnitude of some of the multiple images in the $F814W$-band versus the magnitude predicted by our HFF model. We limit our selection to systems with 26$<mag_{F814W}<$29, and with magnification measurements for at least 2 images in each system, and lower than 15. 
The observed magnitudes are measured from the primary image of each system.  The predicted magnitudes are then obtained by applying the lens model magnifications to a secondary or tertiary image from that system to get a prediction for the observable primary image magnitude. The error bars for the predicted magnitudes are therefore a combination of photometric and model uncertainties.
As one can see, there is a good agreement between the two values, with an average dispersion of 0.3~$mag$.

Our initial mass model of the core of Abell 2744 was more complex, due to an additional mass concentration in the Northern region, close to systems 3, 38, 8, 36, 14 and 37 (Fig.~\ref{multiples}), i.e. in the region in which we corrected the location of multiple images and identified new counter-images.  Testing the need for this additional component we find the resulting RMS to be slightly higher (RMS$=$0.85$\arcsec$) and thus conclude that a third large-scale mass component is not required and not supported by the current observational constraints.  Our hypothesis that this third mass concentration in the model of R14 was only needed to counterbalance the misidentification of System 3, is corroborated by the results of an optimisation run of the pre-HFF model with the identifications for System 3 corrected: again the simple two-component model yields a better $\chi^{2}$ and $RMS$ than the one including a third mass concentration in the cluster core as it is shown in Table~\ref{tablemodel} (first two models), while there are more free parameters.

In order to integrate the mass map within annuli, we choose the location of the overall BCG, i.e., $\alpha=3.586259, \delta=-30.400174$ degrees, as the cluster centre. A circle of radius 45$\arcsec$ (205~kpc) centered on this position
encompasses all multiple images (Fig.~\ref{multiples}). The two-dimensional (cylindrical) mass within this radius is then $M(<200~$kpc$) = (2.162\pm 0.005) \times 10^{14}$\,M$_{\odot}$.

%% file: discussion_hff.tex
\subsection{Comparison with previous mass estimations}
The first strong-gravitational lensing analysis of Abell 2744 performed by \cite{smail97a} using HST/WFPC2 data found a total mass of $M(R<200~{\rm kpc})=(1.85\pm0.32)\times 10^{14}$ M$_{\odot}$. Within the quoted uncertainty this result agrees with our measurement.  Much more recently, a combined strong- plus weak-lensing analysis by \cite{merten11} based on HST/ACS data found a mass for the core component of the cluster of $M(R<250~{\rm kpc})=(2.24\pm0.55)\times 10^{14}$ M$_{\odot}$. Thanks to the deep HFF imaging used in the present work, our analysis reduces the measurement error by an order of magnitude, yielding $M(R<250~{\rm kpc})=(2.765\pm0.008)\times 10^{14}$ M$_{\odot}$. This is the second measurement of a cluster mass with statistical errors of less than 1\%.

As stressed by \cite{jauzac14a}, the precision of mass models from gravitational-lensing studies depends strongly on the mass modeling technique. It is thus noteworthy that the pre-HFF analysis by \cite{lam14}\footnote{\cite{lam14} did not quote a total mass.}, which used a free-form model, predicted the lens-plane position of 18 multiply imaged systems (comprising 55 images) with a mean RMS of ${\sim}0.7\arcsec$. Our parametric model predicts the position of 54 multiply imaged systems (comprising 154 images) with mean RMS=0.79$\arcsec$, a remarkably consistent value for two very different methods.

More recently, \cite{wang15} presented a combined HFF and GLASS analysis of Abell 2744 using 25 (72) multiply-imaged systems (multiple images) amongst the 181 summarized in Table~\ref{multipletable}, selected using an algorithm based on photometric criteria (morphology, spectral energy distribution, ...), and including new spectroscopic redshifts as summarized in Sect.~\ref{slcons}. With this model, they obtain a mass of $M(R<250~{\rm kpc})=(2.43^{+0.04}_{-0.03})\times 10^{14}$ M$_{\odot}$ (A. Hoag private communication).
This value is of the same order of magnitude of what is obtained with our HFF mass model, but it also highlights the fact that at this point, our models are dominated by systematic errors.

\subsection{Magnification measurements discrepancy}
\subsubsection{Comparison between pre-HFF \& HFF CATS models}
The main discrepancy between pre- and post-HFF data models lies in the derived magnification maps.
The following quoted magnification values are sampled from the magnification field at the particular positions, and best-fit redshifts of a small subset of lensed galaxies.
Using a set of 13 multiple images common to pre-HFF and HFF data models, the pre-HFF model of R14 (including the corrected System 3, and thus featuring four cluster-scale halos) provides a median magnification of $3.56 \pm 0.14$, with a position RMS of $0.44\arcsec$. When the N and NW substructures are removed, and therefore only two cluster-scale halos are included, the same pre-HFF model yields a median magnification of $4.69 \pm 0.32$ (RMS${=}0.47\arcsec$) . We use this comparison to obtain an estimate of the systematic uncertainty in the mean magnification due to the model of 0.57. For the exact same set of multiply imaged systems, our HFF model gives a median magnification of $5.61 \pm\,0.10$ (stat) $\pm\, 0.57$ (sys) (RMS${=}0.41\arcsec$). (All these magnification values are also listed in Table~\ref{tablemodel}.)  This discrepancy is not unique to a comparison with the model of R14, but is in fact observed for most pre-HFF models that are publicly available on the \emph{Frontier Fields} lens model page.

While the two pre-HFF models referred to above yield median magnification values that are consistent with each other within 2$\sigma$, they are in clear conflict with the much larger magnification values obtained with the HFF model.  This conflict is unlikely to be caused by the mass components to the N and NW of the cluster core: although, at this stage, these two components are not accurately constrained due to a lack of spectroscopic redshifts for their multiple-image systems, the agreement (within the errors) between the predictions from the 2- and 4-component models of R14 suggests that the impact of the N and NW components on the overall mass distribution is modest.
The significant increase in magnification provided by our high-precision model is equally unlikely to originate from the core region, where the much deeper HFF data have enabled us to correct several misidentifications of multiply imaged systems, and thus to create a simpler mass distribution. Moreover, we have demonstrated that the removal of the third cluster-scale halo, used by R14 to model the cluster core, improved both pre-HFF and HFF mass models (see Table~\ref{tablemodel}). 
Therefore, this third mass component cannot explain the observed discrepancy either.
Finally, the HFF data have allowed us to identify 34 new systems in the core of Abell 2744, providing a highly constrained mass model of this central cluster component.  These new identifications have the strongest impact in the north-western region of the cluster core where previously almost no multiple-image systems had been identified, thus leaving this part of the cluster almost unconstrained and subject to extrapolation from other regions of the core. In Fig.~\ref{multiples} we highlight the pre-HFF multiple-image systems in dark blue; counter-images for systems 3, 8, 14, and 18 as well as several entirely new systems allowed us to map the mass distribution in this region much more accurately (see Fig.\ref{multiples}).
We also extended the region within which multiple-image systems were found in the core farther toward the south. As shown in Fig.~\ref{densprof}, the mass density in this southern region is higher than predicted by the pre-HFF models, resulting in higher magnification in this part of the core.
In conclusion, the plethora of new multiple-image systems discovered in the HFF data has led to significantly tightened lensing constrains in particular across the entire cluster core. The resulting set of constraints span the cluster core more comprehensively. Thereby we have demonstrated that the HFF model presented here is more precise than all previously published models for this cluster.

Regarding the total cluster mass,  we measure $M(R<250~{\rm kpc})=(2.83\pm 0.07)\times 10^{14}$ M$_{\odot}$, using the R14 pre-HFF mass model, in good agreement with the much tighter value of -$M(R<250~{\rm kpc})=(2.765\pm0.008)\times 10^{14}$ M$_{\odot}$ obtained by us in this work from HFF data. Our best-fit mass model provides an improvement of a factor of $\sim$20 in terms of the precision.  We note that although the magnification differs strongly between the two models, the total mass measurements are quite robust and not as dependent on the detailed constraints as magnification measurements are.

\subsubsection{The case of SN HFF14Tom}
\cite{rodney15} presented the discovery of a Type Ia Supernova,
named SN HFF14Tom, at $z=1.3457$, behind Abell 2744, and lensed in the
vicinity of the strong-lensing region. Measurements of the apparent
luminosity distance based on the SNIa light curve provide an estimate
of the magnification for this object of $\mu_{\rm
SN}=2.03\pm0.29$. Comparing this measurement to the predicted
magnifications from a wide range of lens models (including those
presented here), about half of the tested models are within $1\sigma$
of the measured $\mu_{\rm SN}$.  However, they also find evidence for a
mild systematic bias: all models that disagree with the SN
observations are {\it over-predicting} that magnification.  This
includes our HFF mass model, from which we obtain a magnification
$\mu_{\rm HFF, this work}=3.06\pm 0.14$. The model comparisons from
\cite{rodney15} do not isolate a single physical or
methodological source for such a disagreement.  For example, they find
no evidence for a difference between parametric and non-parametric
models, or for those with both strong and weak-lensing constraints vs
those with only strong-lensing constraints.

Using a series of iterations on our baseline model, we have evaluated
two possible avenues for mitigating systematic errors that might
introduce such a bias: (a) increase the number of spectroscopic
redshifts for multiply imaged galaxies, and (b) apply very stringent
cuts when selecting a set of multiple images as strong-lensing
constraints. The model iterations that explore these strategies are:

\begin{description}
\item[(1)] The {\it HFF model} (our fiducial model in this work), which employs the 154 most secured multiple images defined in Sect.~\ref{slcons}, plus all 8 spectroscopic redshifts, including those from \cite{wang15}; 
\item[(2)] The {\it Fewer Spectra} model, using the 154 most secured multiple images but including only 4 of the 8 available spectroscopic redshifts (i.e., using only redshifts that were available in the original HFF
modeling challenge); 
\item[(3)] The {Strict Selection} model, adopting only the 72 multiple images used
by \cite{wang15}, and all 8 spectroscopic redshifts.
\end{description}

\noindent For these three
models we obtain, respectively: 
$\mu_{\rm HFF}=3.06\pm0.14$,
$\mu_{\rm Fewer\,Spectra}=3.42\pm0.15$, 
and $\mu_{\rm Strict\,Selection}=3.07\pm0.13$. 

Comparing the HFF model (1) to the fewer spectra model (2), we can infer
the impact of spectroscopic redshifts (strategy 'a' above).  In this
case we see -- unsurprisingly -- that decreasing the number of
spectroscopic redshifts results in a less accurate prediction for the
SN magnification.  Alternatively, by comparing the HFF model to the
strict selection model (3) we can evaluate the impact of tightening
the criteria for choosing strong-lensing constraints (strategy b).  In
this case we see that limiting the set of multiple images to a "high
quality" subset does not deliver any detectable change in the
magnification prediction.

These tests indicate that the most effective tool for improving model
accuracy is by securing more and better spectroscopic redshifts of
multiply-imaged background sources. Furthermore, the comparison of (1)
to (3) suggests that there may be a limit to the improvement that can be
achieved by tightening the selection of strong-lensing constraints.
That is, once you have a robust set of $\sim$10 spectroscopically
confirmed multiple image sets, adding or subtracting from the
remaining pool of candidate multiple images may have little impact. 

Although this SN HFF14Tom test only samples a single sight line, Figure
3 shows that the HFF model and the strict selection (\emph{Wang+15 systems}, in cyan) model have very
similar radially averaged density profiles.  In particular, both show
a higher density at the edge of the multiply-imaged system region, compared to
pre-HFF mass models.
In a forthcoming paper, we will use weak lensing to better estimate the outskirt density profile.

\subsection{HFF magnification results}
Fig.~\ref{diffampli} summarizes our findings by showing the high-fidelity magnification map from our best-fit HFF mass model for A\,2744, computed for a source at $z_{S}=9$, as well as the surface area in the source plane, $\sigma_{\mu}$, above a given magnification factor. $\sigma_\mu$ is directly proportional to the unlensed comoving volume covered at high redshift at this magnification. The panel b of Fig.~\ref{diffampli} shows $\sigma_\mu$ as a function of $\mu$ for the three models discussed in this paper, as well as for the HFF models using the image set from \cite{wang15}. Following the \cite{wong12} suggestion, we use the area above $\mu = 3$ as a metric to quantify the efficiency of the lensing configuration to magnify high-redshift galaxies and measure $\sigma(\mu > 3) = 0.44$ arcmin$^{2}$ for Abell 2744 with the present HFF model (and $\sigma(\mu > 3) = 0.42$ arcmin$^{2}$ for the HFF model using the set of multiply-imaged systems from \cite{wang15}. Lower values are measured using the pre-HFF models, $\sigma(\mu > 3) = 0.36$ arcmin$^{2}$ for the pre-HFF model with 4 cluster-scale halos, and $\sigma(\mu > 3) = 0.29$ arcmin$^{2}$ for the pre-HFF model with cluster-scale halos. For MACSJ0416, we quote $\sigma(\mu > 3) = 0.26$ arcmin$^{2}$ from \cite{jauzac14a}, almost a factor of two less. Following the trend of the HFF MACSJ0416 strong-lensing results, our present analysis of  Abell 2744 demonstrates the power of HFF data to impressively reduce the statistical errors of both mass and magnification measurements without any changes in the analysis neither the modeling techniques employed.
In the case of Abell 2744, the threefold increase in the number of multiply imaged systems afforded by the exquisite depth and quality of the HFF data improved our estimates of the uncertainty by a factor of $\sim$20 and $\sim$4 for mass and magnification, respectively.

As shown by \cite{atek15}, our high-precision mass model derived from the complete set of HFF data (optical and near-infrared) immediately and significantly improves the constraints on the luminosity function of high-redshift galaxies lensed by this massive lens. 
Similarly, all recent analyses of Abell 2744 and MACSJ0416 based on HFF observations\citep{atek14a,atek15,grillo14,ishigaki15,jauzac14a, jauzac15,lam14,zitrin14} continue to demonstrate and underline the power and legacy value of the HFF data.

%% file: table3.tex
\newpage
\begin {table}
\begin{tabular}{cc}
\, & \,\\
\end{tabular}
\caption{ Multiply imaged systems considered in this work.
$^{+}$ indicate image identifications in which we are less confident, which are not included in the optimization. $^{\ast}$ indicate high-redshift candidate systems identified by \citet{ishigaki15} not included in our model, but for which we assumed $z\sim 8$ for the prediction of the magnification. System \#46 is the high-redhistf system identified by \citet{zitrin14}, and System \#60 is a new system discovered by \citet{wang15}.
However, we include the predicted magnification given by our model. Some of the magnitudes are not quoted because we were facing deblending issues that did not allow us to get reliable measurements. The flux magnification factors come from our best-fit mass model, with errors derived from MCMC sampling.}
\label{multipletable}
\end{table}

\scriptsize

\begin{center}
\par
\tablefirsthead{\hline          \multicolumn{1}{c}{\textbf{ID}} &
                                \multicolumn{1}{c}{\textbf{R.A.}} &
                                \multicolumn{1}{c}{\textbf{Decl.}} &
                                \multicolumn{1}{c}{\textbf{$z_{\rm model}$}} &
                                \multicolumn{1}{c}{m$_{F814W}$} &
                                \multicolumn{1}{c}{$\mu$}
                                \\ \hline }

\tablehead{\hline \multicolumn{5}{l}{\small\sl continued from previous page}\\
                         \hline \multicolumn{1}{c}{\textbf{ID}} &
                                \multicolumn{1}{c}{\textbf{R.A.}} &
                                \multicolumn{1}{c}{\textbf{Decl.}} &
                                \multicolumn{1}{c}{\textbf{$z_{\rm model}$}} &
                                \multicolumn{1}{c}{m$_{F814W}$} &
                                \multicolumn{1}{c}{$\mu$}
                                \\ \hline  }
\tabletail{\hline\multicolumn{5}{r}{\small\sl continued on next page}\\\hline}
\tablelasttail{\hline}
\par
\begin{supertabular}{llllll}
\hline 
 1.1 & 00:14:23.41 & -30:24:14.10 & 1.50 & 26.99$\pm$ 0.16 &  7.4$\pm$ 0.2 \\
 1.2 & 00:14:23.03 & -30:24:24.56 & 1.50 & -- & 11.6$\pm$ 0.4 \\
 1.3 & 00:14:20.69 & -30:24:35.95 & 1.50 & 27.20$\pm$ 0.20 &  4.8$\pm$ 0.1 \\
 2.1 & 00:14:19.98 & -30:24:12.06 & 2.20 & 28.12$\pm$ 0.31 & $>$15 \\
 2.2 & 00:14:23.35 & -30:23:48.21 & 2.20 & 30.00$\pm$ 0.61 &  4.7$\pm$ 0.1 \\
 2.3 & 00:14:20.50 & -30:23:59.63 & 2.20 & -- &  5.8$\pm$ 0.1 \\
 2.4 & 00:14:20.74 & -30:24:07.66 & 2.20 & 27.24$\pm$ 0.24 & 11.1$\pm$ 0.2 \\
 3.1 & 00:14:21.45 & -30:23:37.95 & 3.98 & -- & -- \\
 3.2 & 00:14:21.31 & -30:23:37.69 & 3.98 & 30.61$\pm$ 1.03 & -- \\
 3.3 & 00:14:18.39 & -30:24:06.53 & 3.98 & 27.57$\pm$ 0.09 &  5.5$\pm$ 0.1 \\
 4.1 & 00:14:22.11 & -30:24:09.48 & 3.58 & 27.20$\pm$ 0.21 & 10.4$\pm$ 0.2 \\
 4.2 & 00:14:22.95 & -30:24:05.84 & 3.58 & 26.85$\pm$ 0.09 & 15.0$\pm$ 0.6 \\
 4.3 & 00:14:19.30 & -30:24:32.13 & 3.58 & -- & -- \\
 4.4 & 00:14:22.37 & -30:24:17.69 & 3.58 & 28.94$\pm$ 0.36 & -- \\
 4.5 & 00:14:22.46 & -30:24:18.38 & 3.58 & 28.99$\pm$ 0.33 & $>$15 \\
 5.1 & 00:14:20.02 & -30:23:31.45 & 3.26$\pm$0.06 & 28.18$\pm$ 0.22 & -- \\
 5.2 & 00:14:20.40 & -30:23:28.95 &  & 28.52$\pm$ 0.25 & $>$15 \\
 5.3 & 00:14:19.19 & -30:23:41.14 &  & 29.02$\pm$ 0.19 & $>$15 \\
 6.1 & 00:14:23.65 & -30:24:06.48 & 2.019 & 25.96$\pm$ 0.02 &  5.6$\pm$ 0.1 \\
 6.2 & 00:14:22.57 & -30:24:28.84 & 2.019 & 26.12$\pm$ 0.04 &  5.4$\pm$ 0.1 \\
 6.3 & 00:14:20.74 & -30:24:33.74 & 2.019 & 26.21$\pm$ 0.04 &  6.6$\pm$ 0.1 \\
 7.1 & 00:14:23.58 & -30:24:08.35 & 2.51$\pm$0.07 & 26.72$\pm$ 0.07 &  6.9$\pm$ 0.2 \\
 7.2 & 00:14:22.85 & -30:24:26.73 &  & 26.70$\pm$ 0.11 &  6.3$\pm$ 0.1 \\
 7.3 & 00:14:20.30 & -30:24:35.33 &  & 27.24$\pm$ 0.13 &  5.4$\pm$ 0.1 \\
 8.1 & 00:14:21.53 & -30:23:39.62 & 4.08$\pm$0.16 & 27.74$\pm$ 0.12 & $>$15 \\
 8.2 & 00:14:21.32 & -30:23:39.20 &  & 27.81$\pm$ 0.13 & $>$15 \\
 8.3 & 00:14:18.33 & -30:24:09.23 &  & 28.32$\pm$ 0.18 &  5.4$\pm$ 0.1 \\
 9.1 & 00:14:21.21 & -30:24:18.98 & 2.97$\pm$0.07 & 28.39$\pm$ 0.19 & $>$15 \\
 9.2 & 00:14:20.91 & -30:24:22.47 &  & 28.14$\pm$ 0.25 & $>$15 \\
 9.3 & 00:14:24.04 & -30:23:49.75 &  & 28.78$\pm$ 0.65 &  4.6$\pm$ 0.1 \\
 10.1 & 00:14:21.22 & -30:24:21.16 & 3.61$\pm$0.11 & 27.36$\pm$ 0.16 & $>$15 \\
 10.2 & 00:14:20.97 & -30:24:23.33 &  & 27.11$\pm$ 0.07 & $>$15 \\
 10.3 & 00:14:24.17 & -30:23:49.56 &  & 26.65$\pm$ 0.04 &  4.1$\pm$ 0.1 \\
 11.1 & 00:14:21.93 & -30:24:13.89 & 2.52$\pm$0.10 & 30.26$\pm$ 0.46 &  4.5$\pm$ 0.1 \\
 11.2 & 00:14:23.34 & -30:24:05.23 &  & 29.06$\pm$ 0.36 &  7.5$\pm$ 0.2 \\
 11.3 & 00:14:19.87 & -30:24:32.09 &  & 26.15$\pm$ 0.06 &  5.5$\pm$ 0.1 \\
 11.4 & 00:14:22.69 & -30:24:23.55 &  & 28.85$\pm$ 0.37 &  5.4$\pm$ 0.1 \\
 12.1 & 00:14:22.47 & -30:24:16.09 & 4.52$\pm$0.20 & 28.30$\pm$ 0.23 & -- \\
 12.2 & 00:14:22.38 & -30:24:11.72 &  & 28.00$\pm$ 0.16 & $>$15 \\
 12.3 & 00:14:22.70 & -30:24:10.76 &  & 28.52$\pm$ 0.20 & $>$15 \\
 $^+$12.4 & 00:14:19.10 & -30:24:36.93 &  & -- & 5.5$\pm$0.1 \\
 13.1 & 00:14:22.17 & -30:24:09.21 & 1.60$\pm$0.03 & 29.74$\pm$ 0.52 & $>$15 \\
 13.2 & 00:14:22.51 & -30:24:07.79 &  & 29.14$\pm$ 0.70 & $>$15 \\
 13.3 & 00:14:19.87 & -30:24:28.96 &  & -- &  4.8$\pm$ 0.1 \\
 14.1 & 00:14:21.54 & -30:23:40.69 & 2.47$\pm$0.08 & 27.86$\pm$ 0.19 & $>$15 \\
 14.2 & 00:14:21.23 & -30:23:39.97 &  & 28.96$\pm$ 0.30 &  9.1$\pm$ 0.3 \\
 14.3 & 00:14:18.62 & -30:24:06.14 &  & 29.73$\pm$ 1.38 &  5.5$\pm$ 0.1 \\
 18.1 & 00:14:21.78 & -30:23:44.02 & 5.66 & 26.73$\pm$ 0.10 & $>$15 \\
 18.2 & 00:14:21.21 & -30:23:44.29 & 5.66 & -- &  3.1$\pm$ 0.1 \\
 18.3 & 00:14:18.27 & -30:24:16.11 & 5.66 & 27.63$\pm$ 0.15 &  4.6$\pm$ 0.1 \\
 $^+$19.1 & 00:14:21.34 & -30:23:50.78 &  & -- & 8.0$\pm$0.2 \\
 19.2 & 00:14:21.94 & -30:23:48.07 & 2.21$\pm$0.06 & 29.16$\pm$ 0.41 & 12.2$\pm$ 0.6 \\
 19.3 & 00:14:18.89 & -30:24:14.53 &  & -- &  8.0$\pm$ 0.2 \\
 20.1 & 00:14:23.10 & -30:24:10.68 & 2.79$\pm$0.07 & 28.65$\pm$ 0.18 & 14.3$\pm$ 0.5 \\
 20.2 & 00:14:22.84 & -30:24:19.69 &  & 29.39$\pm$ 0.29 &  9.8$\pm$ 0.2 \\
 20.3 & 00:14:19.68 & -30:24:34.44 &  & -- &  5.2$\pm$ 0.1 \\
 21.1 & 00:14:23.08 & -30:24:11.23 & 2.72$\pm$0.08 & 29.05$\pm$ 0.24 & $>$15 \\
 21.2 & 00:14:22.86 & -30:24:19.14 &  & 28.81$\pm$ 0.21 & 12.9$\pm$ 0.3 \\
 21.3 & 00:14:19.67 & -30:24:34.62 &  & 31.34$\pm$ 1.31 &  5.1$\pm$ 0.1 \\
 22.1 & 00:14:21.10 & -30:24:41.80 & 2.91$\pm$0.09 & 28.24$\pm$ 0.19 &  8.0$\pm$ 0.2 \\
 22.2 & 00:14:24.02 & -30:24:15.90 &  & 28.78$\pm$ 0.24 &  7.1$\pm$ 0.2 \\
 22.3 & 00:14:23.17 & -30:24:32.51 &  & 28.28$\pm$ 0.26 &  8.7$\pm$ 0.3 \\
 23.1 & 00:14:21.16 & -30:24:37.99 & 2.96$\pm$0.11 & 28.12$\pm$ 0.16 &  9.3$\pm$ 0.3 \\
 23.2 & 00:14:22.45 & -30:24:34.99 &  & 28.09$\pm$ 0.19 & 12.1$\pm$ 0.6 \\
 23.3 & 00:14:24.13 & -30:24:06.64 &  & 29.31$\pm$ 0.48 &  4.9$\pm$ 0.1 \\
 24.1 & 00:14:23.02 & -30:24:16.14 & 1.17$\pm$0.02 & 27.37$\pm$ 0.13 & 12.1$\pm$ 0.4 \\
 24.2 & 00:14:22.83 & -30:24:21.36 &  & 30.29$\pm$ 1.06 &  6.7$\pm$ 0.1 \\
 24.3 & 00:14:20.96 & -30:24:32.77 &  & 28.65$\pm$ 0.27 &  5.3$\pm$ 0.1 \\
 25.1 & 00:14:22.67 & -30:24:09.87 & 1.33$\pm$0.03 & 28.67$\pm$ 0.22 & 13.4$\pm$ 0.5 \\
 25.2 & 00:14:22.12 & -30:24:12.00 &  & 28.85$\pm$ 0.22 &  9.6$\pm$ 0.2 \\
 25.3 & 00:14:20.21 & -30:24:29.86 &  & 29.07$\pm$ 0.23 &  4.7$\pm$ 0.1 \\
 26.1 & 00:14:22.55 & -30:24:34.87 & 2.31$\pm$0.05 & 27.95$\pm$ 0.12 &  9.8$\pm$ 0.4 \\
 26.2 & 00:14:21.69 & -30:24:38.11 &  & 26.78$\pm$ 0.10 & $>$15 \\
 26.3 & 00:14:24.02 & -30:24:10.69 &  & 28.82$\pm$ 0.23 &  5.2$\pm$ 0.1 \\
 27.1 & 00:14:19.38 & -30:24:11.34 & 2.52$\pm$0.07 & 29.25$\pm$ 0.26 &  9.8$\pm$ 0.3 \\
 27.2 & 00:14:22.97 & -30:23:46.23 &  & 30.35$\pm$ 0.71 &  5.6$\pm$ 0.1 \\
 27.3 & 00:14:20.52 & -30:23:51.55 &  & 29.81$\pm$ 0.39 &  5.9$\pm$ 0.1 \\
 28.1 & 00:14:19.31 & -30:24:18.24 & 5.72$\pm$0.34 & 30.92$\pm$ 1.60 &  7.6$\pm$ 0.2 \\
 28.2 & 00:14:23.48 & -30:23:45.45 &  & -- &  5.3$\pm$ 0.1 \\
 28.3 & 00:14:20.47 & -30:23:52.69 &  & -- &  6.4$\pm$ 0.1 \\
 28.4 & 00:14:20.99 & -30:24:04.94 &  & 32.16$\pm$ 4.80 &  4.4$\pm$ 0.1 \\
 29.1 & 00:14:19.76 & -30:23:51.52 & 3.13$\pm$0.18 & 30.10$\pm$ 0.64 & -- \\
 29.2 & 00:14:19.33 & -30:24:01.54 &  & 29.31$\pm$ 0.32 & $>$15 \\
 30.1 & 00:14:21.84 & -30:23:50.80 & 1.18$\pm$0.01 & 27.65$\pm$ 0.12 &  9.2$\pm$ 0.3 \\
 30.2 & 00:14:20.81 & -30:23:53.47 &  & 27.25$\pm$ 0.06 &  9.3$\pm$ 0.2 \\
 30.3 & 00:14:19.66 & -30:24:06.13 &  & 27.64$\pm$ 0.13 &  7.2$\pm$ 0.2 \\
 31.1 & 00:14:20.62 & -30:24:11.40 & 4.18$\pm$0.38 & 29.21$\pm$ 0.46 & $>$15 \\
 31.2 & 00:14:20.09 & -30:24:14.82 &  & 29.43$\pm$ 0.40 & $>$15 \\
 $^+$31.3 & 00:14:23.96 & -30:23:43.88 &  & -- & 3.9$\pm$0.1 \\
 32.1 & 00:14:20.06 & -30:24:16.98 & 4.16$\pm$0.22 & 28.90$\pm$ 0.23 & $>$15 \\
 32.2 & 00:14:20.80 & -30:24:12.07 &  & 28.58$\pm$ 0.25 & 13.6$\pm$ 0.2 \\
 $^+$32.3 & 00:14:23.95 & -30:23:45.53 &  & -- & 4.2$\pm$0.1 \\
 33.1 & 00:14:20.33 & -30:24:11.33 & 5.7 & -- & 12.3$\pm$ 0.2 \\
 33.2 & 00:14:20.26 & -30:24:12.20 & 5.7 & 28.05$\pm$ 0.11 & -- \\
 33.3 & 00:14:24.16 & -30:23:43.53 & 5.7 & 29.32$\pm$ 0.32 &  4.5$\pm$ 0.1 \\
 34.1 & 00:14:22.42 & -30:24:39.03 & 2.53$\pm$0.04 & 29.51$\pm$ 0.46 & -- \\
 34.2 & 00:14:22.52 & -30:24:38.61 &  & 29.25$\pm$ 0.29 & $>$15 \\
 34.3 & 00:14:24.14 & -30:24:16.32 &  & 27.80$\pm$ 0.14 &  6.3$\pm$ 0.1 \\
 36.1 & 00:14:21.53 & -30:23:39.99 & 2.53$\pm$0.06 & 28.73$\pm$ 0.32 & $>$15 \\
 36.2 & 00:14:21.28 & -30:23:39.48 &  & 29.07$\pm$ 0.36 & $>$15 \\
 36.3 & 00:14:18.60 & -30:24:05.39 &  & 29.16$\pm$ 0.43 &  5.5$\pm$ 0.1 \\
 37.1 & 00:14:21.37 & -30:23:41.69 &  & 28.41$\pm$ 0.17 & $>$15 \\
 37.2 & 00:14:21.29 & -30:23:41.47 &  & -- & -- \\
 37.3 & 00:14:19.07 & -30:24:01.07 &  & 30.34$\pm$ 1.24 & 10.2$\pm$ 0.3 \\
 38.1 & 00:14:21.46 & -30:23:38.78 & 4.36$\pm$0.09 & 29.31$\pm$ 0.19 & $>$15 \\
 38.2 & 00:14:21.35 & -30:23:38.56 &  & 28.49$\pm$ 0.17 & $>$15 \\
 38.3 & 00:14:18.33 & -30:24:07.67 &  & 29.73$\pm$ 0.48 &  5.7$\pm$ 0.1 \\
 39.1 & 00:14:21.31 & -30:23:33.11 & 3.37$\pm$0.04 & 31.59$\pm$ 1.61 & -- \\
 39.2 & 00:14:21.25 & -30:23:33.03 &  & 27.72$\pm$ 0.11 & -- \\
 39.3 & 00:14:18.61 & -30:23:57.74 &  & 28.57$\pm$ 0.14 &  8.8$\pm$ 0.3 \\
 40.1 & 00:14:21.38 & -30:23:33.59 & 2.60$\pm$0.29 & 28.52$\pm$ 0.22 & $>$15 \\
 40.2 & 00:14:21.17 & -30:23:33.19 &  & 28.93$\pm$ 0.24 & -- \\
 41.1 & 00:14:23.80 & -30:23:58.50 & 4.06$\pm$0.19 & 30.09$\pm$ 0.57 &  5.4$\pm$ 0.1 \\
 41.2 & 00:14:22.45 & -30:24:27.97 &  & 29.04$\pm$ 0.61 &  4.4$\pm$ 0.1 \\
 41.3 & 00:14:20.03 & -30:24:30.60 &  & 29.71$\pm$ 0.43 &  7.2$\pm$ 0.2 \\
 41.4 & 00:14:21.70 & -30:24:14.55 &  & 29.65$\pm$ 0.32 &  5.3$\pm$ 0.1 \\
 42.1 & 00:14:23.35 & -30:24:02.19 & 3.61$\pm$0.14 & 28.29$\pm$ 0.13 &  9.5$\pm$ 0.3 \\
 $^+$42.2 & 00:14:21.83 & -30:24:11.74 &  & -- & 4.7$\pm$0.1 \\
 42.3 & 00:14:19.58 & -30:24:31.08 & 3.61$\pm$0.14 & 30.74$\pm$ 1.11 &  6.4$\pm$ 0.2 \\
 42.4 & 00:14:22.62 & -30:24:23.00 &  & 28.27$\pm$ 0.31 &  5.0$\pm$ 0.1 \\
 43.1 & 00:14:23.48 & -30:24:09.01 & 2.67$\pm$0.09 & 30.83$\pm$ 2.13 &  7.8$\pm$ 0.2 \\
 43.2 & 00:14:20.15 & -30:24:35.34 &  & -- &  5.6$\pm$ 0.1 \\
 44.1 & 00:14:20.03 & -30:24:25.08 & 1.81$\pm$0.06 & 30.68$\pm$ 1.30 &  6.3$\pm$ 0.1 \\
 44.2 & 00:14:23.21 & -30:23:59.15 &  & 31.82$\pm$ 4.49 &  6.0$\pm$ 0.1 \\
 45.1 & 00:14:20.36 & -30:23:54.48 & 3.31$\pm$0.21 & 28.46$\pm$ 0.17 &  6.0$\pm$ 0.1 \\
 45.2 & 00:14:19.54 & -30:24:14.21 &  & 28.08$\pm$ 0.16 & 10.2$\pm$ 0.3 \\
 45.3 & 00:14:20.85 & -30:24:04.66 &  & -- &  5.4$\pm$ 0.1 \\
 $^+$45.4 & 00:14:23.38 & -30:23:46.13 &  & -- & 8.8$\pm$0.3 \\
 46.1 & 00:14:22.81 & -30:24:02.71 & 10.0 & -- & $>$15 \\
 46.2 & 00:14:22.20 & -30:24:05.35 & 10.0 & -- & $>$15 \\
 46.3 & 00:14:18.60 & -30:24:31.32 & 10.0 & -- &  4.7$\pm$ 0.1 \\
 47.1 & 00:14:20.75 & -30:23:31.65 & 3.69$\pm$0.65 & 27.81$\pm$ 0.15 & -- \\
 47.2 & 00:14:20.60 & -30:23:32.07 &  & 27.97$\pm$ 0.11 & -- \\
 $^+$47.3 & 00:14:21.52 & -30:23:31.71 &  & -- & $>$156 \\
 $^+$47.4 & 00:14:18.80 & -30:23:53.28 &  & -- & 10.8$\pm$0.4 \\
 48.1 & 00:14:22.62 & -30:24:10.23 & 1.62$\pm$0.23 & -- & $>$15 \\
 48.2 & 00:14:22.27 & -30:24:11.28 &  & 30.71$\pm$ 1.14 & $>$15 \\
 $^+$48.3 & 00:14:19.70 & -30:24:30.90 &  & -- & 4.5$\pm$0.1 \\
 49.1 & 00:14:22.24 & -30:24:29.78 & 1.24$\pm$0.04 & 29.05$\pm$ 0.50 & 11.1$\pm$ 0.4 \\
 49.2 & 00:14:21.65 & -30:24:31.70 &  & 30.11$\pm$ 0.70 & $>$15 \\
 $^+$49.3 & 00:14:23.40 & -30:24:11.30 &  & -- & 5.5$\pm$0.1 \\
 50.1 & 00:14:18.71 & -30:24:05.82 & 4.35$\pm$0.25 & 29.21$\pm$ 0.36 &  6.9$\pm$ 0.2 \\
 50.2 & 00:14:22.55 & -30:23:39.48 &  & 30.02$\pm$ 0.55 &  9.4$\pm$ 0.4 \\
 $^+$50.3 & 00:14:20.43 & -30:23:37.44 &  & -- & 2.5$\pm$0.1 \\
 51.1 & 00:14:20.84 & -30:24:19.97 & 3.88$\pm$1.88 & 28.11$\pm$ 0.18 & -- \\
 51.2 & 00:14:20.75 & -30:24:20.40 &  & 28.48$\pm$ 0.15 & $>$15 \\
 52.1 & 00:14:20.78 & -30:23:49.23 &  & -- & $>$15 \\
 52.2 & 00:14:20.67 & -30:23:49.68 &  & 28.94$\pm$ 0.60 & $>$15 \\
 53.1 & 00:14:19.16 & -30:24:05.73 & 5.15$\pm$0.38 & -- & 11.1$\pm$ 0.4 \\
 53.2 & 00:14:20.05 & -30:23:48.13 &  & 29.64$\pm$ 0.52 & 14.7$\pm$ 0.5 \\
 54.1 & 00:14:22.17 & -30:24:35.61 & 3.20$\pm$0.15 & 30.46$\pm$ 0.77 & $>$15 \\
 54.2 & 00:14:21.20 & -30:24:37.19 &  & 29.71$\pm$ 0.74 & 14.2$\pm$ 0.6 \\
 $^+$54.3 & 00:14:24.23 & -30:24:03.02 &  & -- & 4.5$\pm$0.1 \\
 54.4 & 00:14:21.62 & -30:24:36.99 & 3.20$\pm$0.15 & -- & $>$15 \\
 55.1 & 00:14:23.29 & -30:24:17.09 & 1.50 & 27.32$\pm$ 0.13 & 13.6$\pm$ 0.5 \\
 55.2 & 00:14:23.13 & -30:24:22.18 & 1.50 & 27.44$\pm$ 0.09 &  8.9$\pm$ 0.2 \\
 55.3 & 00:14:20.57 & -30:24:36.31 & 1.50 & 27.37$\pm$ 0.15 &  4.6$\pm$ 0.1 \\
 56.1 & 00:14:19.81 & -30:24:08.24 & 2.20 & 28.35$\pm$ 0.26 & $>$15 \\
 56.2 & 00:14:20.69 & -30:24:03.06 & 2.20 & -- &  4.7$\pm$ 0.1 \\
 56.3 & 00:14:20.27 & -30:23:57.45 & 2.20 & -- &  5.8$\pm$ 0.1 \\
 56.4 & 00:14:23.21 & -30:23:46.67 & 2.20 & 28.83$\pm$ 0.47 &  4.9$\pm$ 0.1 \\
 $^{\ast}$57.1 & 00:14:23.68 & -30:24:17.68 &  & -- & $>$15 \\
 $^{\ast}$57.2 & 00:14:20.89 & -30:24:40.54 &  & -- & 8.5$\pm$0.3 \\
 $^{\ast}$57.3 & 00:14:23.24 & -30:24:28.19 &  & -- & 6.6$\pm$0.2 \\
 $^{\ast}$58.1 & 00:14:18.74 & -30:23:58.70 &  & -- & 12.6$\pm$0.5 \\
 $^{\ast}$58.2 & 00:14:21.42 & -30:23:39.98 &  & -- & $>$15 \\
 59.1 & 00:14:20.23 & -30:24:32.15 & 2.41$\pm$0.11 & 27.87$\pm$ 0.10 &  6.5$\pm$ 0.2 \\
 59.2 & 00:14:23.55 & -30:24:03.53 &  & 28.45$\pm$ 0.17 &  6.0$\pm$ 0.1 \\
 60.1 & 00:14:23.54 & -30:24:14.37 & 1.55$\pm$0.04 & 28.55$\pm$ 0.23 &  6.7$\pm$ 0.1 \\
 60.2 & 00:14:22.97 & -30:24:27.18 &  & 29.05$\pm$ 0.34 & 10.3$\pm$ 0.4 \\
 60.3 & 00:14:20.97 & -30:24:36.54 &  & 30.33$\pm$ 1.38 &  5.5$\pm$ 0.1 \\
  $^+$61.1 & 00:14:22.89 & -30:24:13.62 & 1.2$\pm$0.2 & 31.22$\pm$1.86 & 11.0$\pm$2.6 \\
  $^+$61.2 & 00:14:22.86 & -30:24:16.02 & & 29.48$\pm$0.31 & $>$15 \\
\hline
\end{supertabular}
\end{center}